# Modulation Doping of Silicon using Aluminium-induced Acceptor States in Silicon Dioxide

Dirk König[1,2,3], Daniel Hiller[2,3], Sebastian Gutsch[3], Margit Zacharias[3] & Sean Smith[1]

[1] *Integrated Materials Design Centre (IMDC), UNSW, Sydney, Australia*

[2] *School of Photovoltaic and Renewable Energy Engineering (SPREE), UNSW, Sydney, Australia*

[3] *Laboratory for Nanotechnology, Dept. of Microsystems Engineering (IMTEK), University of Freiburg, Germany*

Correspondence should be addressed to: D. K. (solidstatedirk@gmail.com) or D. H. (daniel.hiller@imtek.uni-freiburg.de)

All electronic, optoelectronic or photovoltaic applications of silicon depend on controlling majority charge carriers via doping with impurity atoms. Nanoscale silicon is omnipresent in fundamental research (quantum dots, nanowires) but also approached in future technology nodes of the microelectronics industry. In general, silicon nanovolumes, irrespective of their intended purpose, suffer from effects that impede conventional doping due to fundamental physical principles such as out-diffusion, statistics of small numbers, quantum- or dielectric confinement. In analogy to the concept of modulation doping, originally invented for III-V semiconductors, we demonstrate a heterostructure modulation doping method for silicon. Our approach utilizes a specific acceptor state of aluminium atoms in silicon dioxide to generate holes as majority carriers in adjacent silicon. By relocating the dopants from silicon to silicon dioxide, Si nanoscale doping problems are circumvented. In addition, the concept of aluminium-induced acceptor states for passivating hole selective tunnelling contacts as required for high-efficiency photovoltaics is presented and corroborated by first carrier lifetime measurements.

## Introduction

Conventional silicon doping is increasingly impeded due to the spatial dimensions approached by nanotechnology. Several fundamental physical principles counteract the substitutional incorporation of dopant impurities (e.g. B, P, or As) on Si lattice sites and their ionization to become electronically active donors of majority charge carriers. The formation energy for the substitutional dopant integration[1,2] as well as the ionization energy[3-6] increase strongly with decreasing Si dimensions. Moreover, conventional doping of Si nanoscale devices faces severe technological challenges: diffusion-related, steep radial gradients in the doping profile of Si nanowires[7-9]; the inadvertent but inevitable diffusion of source/drain dopants into field-effect transistor (FET)-channels[10]; surface segregation and inactivation of dopants[11]; and statistical fluctuations by random numbers/positions of dopants in Si nanovolumes[12-14]. These problems render conventional Si doping unsuitable for future nanoelectronics.

Here, we demonstrate a novel approach: modulation doping of Si by aluminium (Al)-induced acceptor states in silicon dioxide ($SiO_2$). Using computer-aided materials design, we predict that Al atoms in $SiO_2$ generate acceptor states 0.8 eV below the Si valence band edge. These states capture electrons from Si over a distance of several nanometres, providing holes to Si as majority charge carriers. We confirm this concept experimentally via capacitance-voltage (CV) and deep level transient spectroscopy (DLTS) on $SiO_2$:Al/Si-based MOS capacitors. This doping technique is robust against out-diffusion, quantum confinement, dielectric confinement, and self-purification since it relocates doping from confined Si nanovolumes to adjacent bulk-like $SiO_2$.

## Results

**Density functional theory simulations.** Acceptor modulation doping of $SiO_2$ was modelled via real-space density functional theory (DFT). We embedded a $Si_{10}$ nanocrystal[15] in three monolayers (MLs) of $SiO_2$, presenting the ultimate theoretical test of the doping concept because Si nanocrystal, $SiO_2$ and modulation acceptor (Al) form one approximant. Figures 1a and b show the highest occupied (HO) molecular orbital (MO) and the lowest unoccupied (LU) MO associated with Al. High probability densities occur within the $Si_{10}$ nanocrystal and at the Al acceptor, although both are separated by 3 ML of perfect $SiO_2$. This finding shows that the $Si_{10}$ nanocrystal can still be positively ionised despite significant quantum confinement-induced bandgap widening. Fig. 1c shows the electronic density of states (DOS) of $Si_{10}$ nanocrystals in pure $SiO_2$ vs. $SiO_2$:Al, along with the HOMO and LUMO of a 1.9 nm Si nanocrystal, fully terminated with hydroxyl (OH) groups, featuring $SiO_2$ embedding[16]. The Al acceptor state ($\beta$-LUMO) exists 0.26 eV below the HOMO of the 1.9 nm Si nanocrystal, clearly showing that Si nanocrystals of $d \geq 1.9$ nm can obtain positive majority charge carriers (holes) from Al acceptors in $SiO_2$. Transferring the situation to bulk Si, the Al acceptor state is located ~0.8 eV below the valence band edge. The principle of direct acceptor modulation doping is shown in the electronic band structures in Fig. 1d for Si nanocrystals and in Fig. 1e for bulk Si.

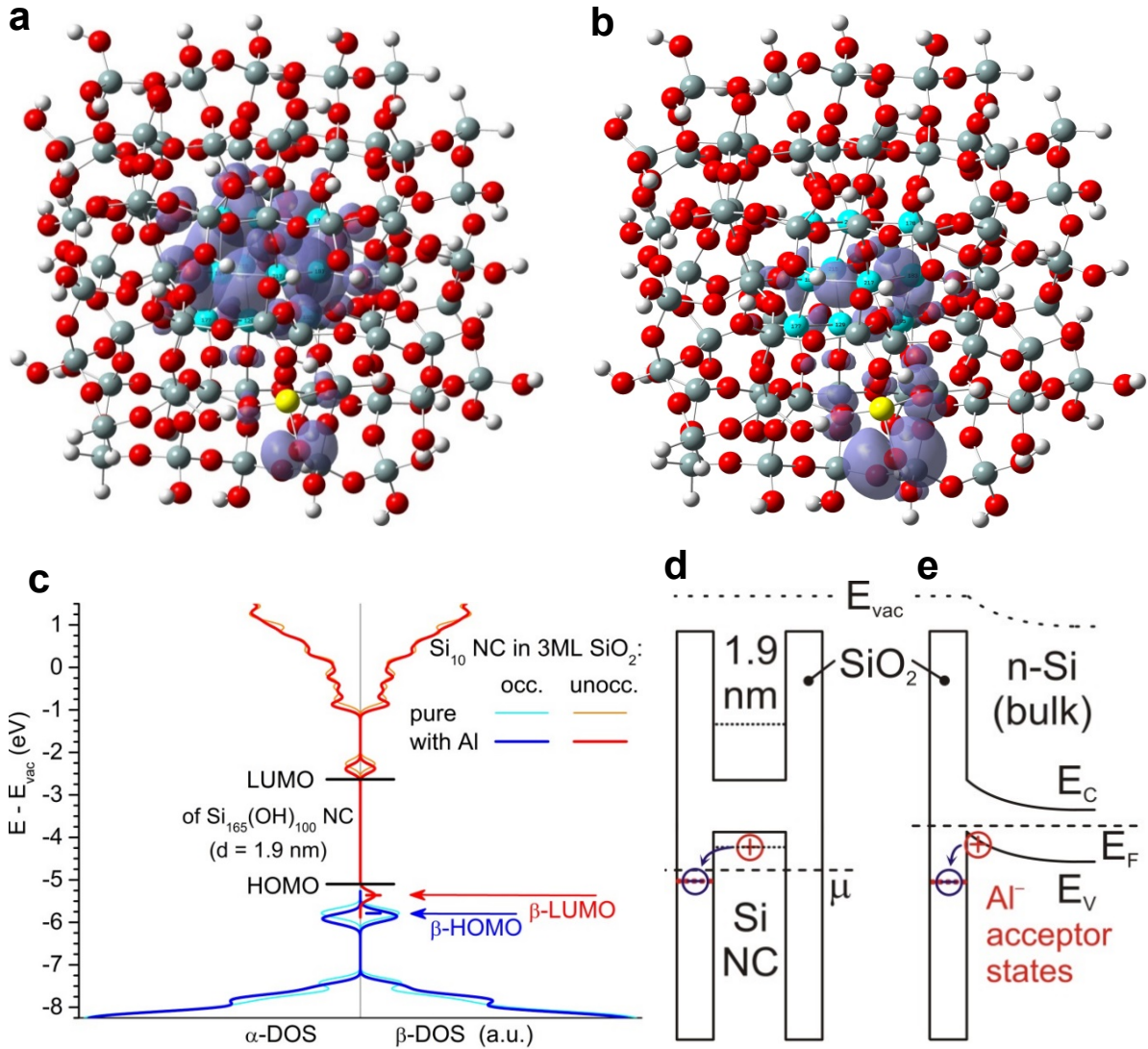

**Figure 1: SiO$_2$ modulation doping with Al acceptors, DFT results. a,b,** Si$_{10}$ nanocrystal (cyan) in three ML SiO$_2$ (O is red, Si grey, H white) with Al atom (yellow) replacing Si in outermost SiO$_2$ shell, showing β-HOMO (**a**) and β-LUMO (**b**) as iso-density plots of 4×10$^{-4}$ *e*/cubic Bohr radius. **c** Electronic DOS (blue, red) of that approximant (energy scale refers to vacuum level E$_{vac}$). DOS for pure SiO$_2$ embedding (cyan, orange) is shown for comparison along with the HOMO and LUMO of a 1.9 nm Si nanocrystal fully terminated with OH groups[15]. The two possible MO spin orientations are noted by α and β due to unpaired electron configuration caused by the acceptor (doublet). **d,** Band structures showing the principle of direct modulation doping for a Si nanocrystal (NC) in SiO$_2$ and **e,** for Si bulk terminated with SiO$_2$:Al layer. Occupation of electronic states are described by chemical potential μ and Fermi level E$_F$, respectively.

**Electrical characterisation of modulation doped silicon.** Figure 2a shows a schematic of the Si/SiO$_2$:Al MOS capacitor, fabricated via plasma enhanced chemical vapour deposition (PECVD) of SiO$_2$ on n-type Si wafers combined with thermal atomic layer deposition (ALD) of 1 or 2 Al–O MLs. After deposition, the structures were annealed for 30 sec at 1000°C in Ar ambient; subsequently, thermally evaporated aluminium contacts were structured. As evident from Fig. 1e, successful modulation doping of Si will manifest itself as a fixed negative charge in the SiO$_2$:Al film. Using 1 MHz high-frequency capacitance-voltage (CV), we measured a change in the fixed charge of $\Delta Q_{fix}$ = −2.3×10$^{12}$ cm$^{-2}$ for 1 ALD Al–O ML in SiO$_2$ (Al1) compared with pure SiO$_2$ reference samples (Al0) (Fig. 2b). The flatband voltage $V_{FB}$ was +0.9 V and +1.9 V for samples Al1 and Al2, respectively, whereas for the reference sample Al0, the flatband condition was achieved for -0.2 V. By using n-type Si with a donor density of 3.5×10$^{15}$ cm$^{-3}$ and Al as the gate metal, we keep the work function difference between the gate and n-Si negligible (-0.004 eV). High-frequency CV does not allow for a contribution of Si/SiO$_2$ interface states to the CV signal[17], which can therefore be ruled out as having an effect on the CV curve.

High energy resolution deep level transient spectroscopy (HERA-DLTS) was carried out to further characterise the Al-induced acceptor state. Figure 2d plots the transient capacitance over the reverse bias $V_R$ at 102 K. Clearly, sample Al1 shows a dominant signal at $V_R$ = −3.5 V, whereas no transient capacitance is detected for the reference sample. The peak originates from electrons escaping from the Al-state into the Si wafer via direct tunnelling (scattering, trap-assisted tunnelling and hopping are suppressed at low temperatures). The transient time constant $T_W$ = 31 ms was kept comparatively short at the cost of signal intensity to further eliminate any trap-assisted tunnelling that is less dynamic. Using the Poisson equation, the energetic position of the Al-acceptor state in the SiO$_2$ bandgap was calculated to be 0.5 eV below the Si valence band edge. This energy value underlines the accuracy of DFT calculations, yielding 0.8 eV. The full with half maximum (FWHM) of $V_R$ was $\Delta V_R$ = 0.66 V, cf. Fig. 2c. The FWHM translates into a vertical-spatial distribution of Al atoms of ca. 0.27 nm, accounting for RTO interface roughness and binding of Al to the RTO, as well as binding to the CVD capping oxide.

This finding confirms our h-DFT results, where Al in SiO$_2$ provides an acceptor state even for small Si NCs and corroborates that Al undergoes virtually no diffusion during the activation anneal, in accordance with its diminutive diffusion coefficient[18]. In Fig. 2e, we demonstrate the reverse pulse scheme to sample electron capture into Al acceptors and plot transient capacitance over the pulse time $t_P$ at 502 K. The modulation doped sample Al1 showed one broad peak at $t_P$ = 2.0 s, with a maximum transient signal of 146 nF/cm$^2$ or 9.11×10$^{11}$ cm$^{-2}$ transferred charge, accounting for ca. 40 % of all negative Al acceptors. No electron capture signal was detected for the reference sample. Time constants $T_W$ and $t_P$ were chosen to be large compared to low-temperature electron escape in order to account for the maximum density of charged Al acceptors. The substantially higher two-peak signal of sample Al2 shows that we control the density of active modulation acceptors in $SiO_2$, which was confirmed by HF-CV results.

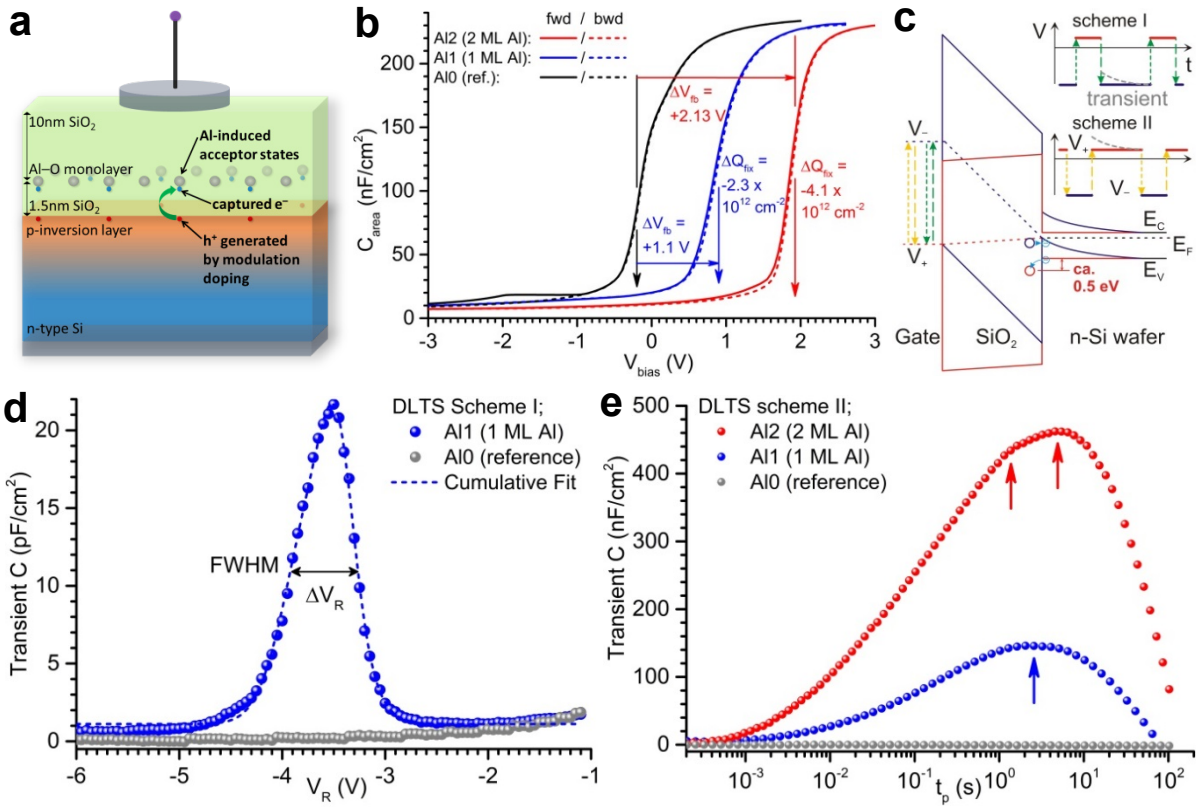

**Figure 2: $SiO_2$ modulation doping with Al acceptors using Al/$SiO_2$/Al-O/$SiO_2$/Si MOS structures, electronic characterisation. a,** Sample structure showing Al modulation acceptors charged from Si substrate by electron tunnelling. **b,** CV curves of reference sample Al0 (no Al-O), Al1 (1 ML Al-O) and Al2 (2 ML Al-O) measured at T = 300 K; $\Delta V_{fb}$ and $\Delta Q_{fix}$ due to negatively ionised Al shown by coloured arrows. **c,** Band structure scheme of charge transient measurements for electron release [scheme I] and for electron capture [scheme II] by Al in DLTS. **d,** Electron release of Al in $SiO_2$ with pulse time $t_p$ = 205 µs, transient time $T_W$ = 31 ms and pulse voltage $V_p$ = +0.5 V as function of reference voltage $V_R$ measured at T = 102 K to freeze out inelastic scattering and trap-assisted processes for maximum energy resolution by direct electron tunnelling into Si. **e,** Electron capture with transient time $T_W$ = 3.63 s, pulse voltage $V_p$ = -4 V and reference voltage $V_R$ = 0 V as function of pulse time $t_p$ measured at T = 502 K to activate all transport paths (hopping, direct and trap assisted tunnelling) for maximum occupation probability of Al in $SiO_2$. Arrows show sub-peaks in accord with Al-O MLs. Capacitance per area scale changes from pF/cm$^2$ in (d) to nF/cm$^2$ in (e).

**Field effect passivation by Al-induced acceptor states.** First results on 1 carrier lifetime measurements carried out by microwave-photoconduction decay (µW-PCD)[19,20] are shown in Fig. 3 using mechanical grade double side polished 525 µm thick Czochralski Si wafers with an ultrathin rapid-thermal oxide. Samples were not passivated in hydrogen ($H_2$) or forming gas (FG). The μW-PCD spectra show a minority (hole) carrier lifetime $\tau_{hole}$ = 1 ms at an excess majority carrier density corresponding to 1 sun illumination ($\Delta p = 10^{15}$ cm$^{-3}$), presenting an increase over the reference sample with undoped annealed CVD-SiO$_2$ by a factor of 100. The corresponding emitter saturation current density[21] is $j_0 = 1.05 \times 10^{-13}$ A/cm$^2$. This extremely low value obtained on preliminary samples shows the potential of SiO$_2$ modulation doping for high efficiency silicon solar cells.

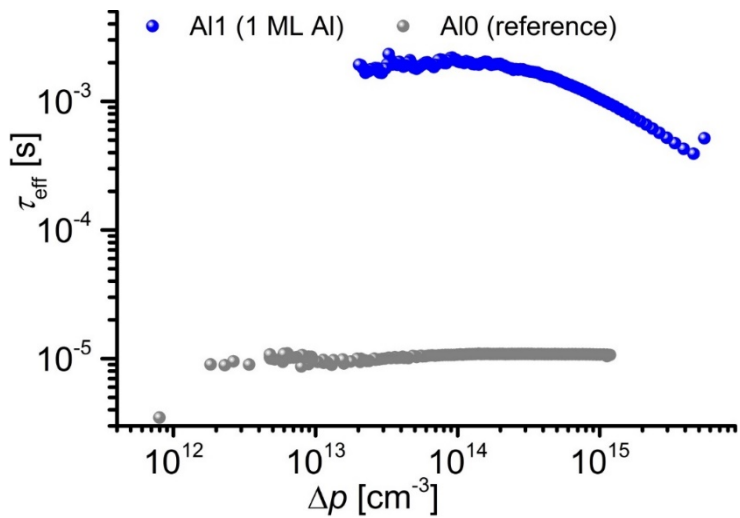

**Figure 3: Effect of SiO$_2$:Al modulation doping on effective minority carrier (hole) lifetime.** Double-side polished 1.6 Ωcm phosphorus-doped 525 μm Czochralski Si wafers with RTO/Al–O/SiO$_2$ stacks on both sides (sample Al1) show more than 2 orders of magnitude higher lifetimes compared to the RTO/SiO$_2$ reference sample (Al0). The effective minority carrier lifetime of Al1 at an excess minority carrier concentration corresponding to 1 sun illumination ($\Delta p = 10^{15}$ cm$^{-3}$) is $\tau_{hole}$ = 1 ms.

**Discussion**

Modulation doping of III-V semiconductors was discovered in the late 1970s[22]. Homostructure modulation doping was proposed for Si and Ge nanowires using conventional dopants in a Si shell around nanowires, allowing for separation of majority charge carriers from their parent donor impurities[23,24]. However, this approach does not solve issues of dopant inter-diffusion, deficient dopant ionisation or statistical distributions of dopant number and position. An impressive work-around, referring back to junctionless FETs (Lilienfeld, 1925), was demonstrated by Colinge et al.[25]. This approach solves the problem of highly abrupt p-n junctions, though ultimately scaled devices again suffer from nanoscale-Si doping obstacles. Hypervalent doping of free-standing Si quantum dots was demonstrated using chemical surface engineering[26]. However, so far this approach is not at all CMOS compatible. Molecular monolayer doping[27,28] was developed to achieve ultra-shallow p-n junctions via RTA-diffusion. Since this concept is based on classical impurity dopants, its applicability in low-dimensional systems is subject to the same constraints outlined above.

Heterostructure modulation doping of Si using intentionally designed impurities in $SiO_2$ presents a true paradigm shift by "outsourcing" dopants to the surrounding matrix. This approach can acceptor-dope Si MOS-FETs by incorporating Al atoms in the insulating $SiO_2$ trench or into the base and coating for fin-FETs, which circumvents all mentioned nanoscale doping problems. As shown exemplarily in Figures 4a and c, the source and drain areas of an intrinsic Si fin-FET become p-type conductive by incorporating Al-atoms in the buried oxide (BOX). The absence of dopants within the active Si volume eliminates dopant impurity scattering, resulting in lower heat generation, higher carrier mobilities and consequently lower operating voltages. These features are very beneficial for advancing ultra-large scale integration (ULSI) and ultra low power applications. Moreover, modulation doping avoids the so-called dopant fingerprint of nanoscale MOS-devices, i.e. statistical performance fluctuations due to variations in the exact number and distribution of dopants[4,12,14]. Coulomb repulsion between the charged Al induced acceptor states self-regulates the amount of holes generated and prevents that all such states capture an electron. In fact, approximately only one in a hundred acceptor states (1 Al-O monolayer) can be charged and hole-doping is easily controlled via 1 the areal coverage of the Si nanovolume with $SiO_2$:Al.

Apart from microelectronics, Si-modulation doping can enhance passivating tunnelling contacts in heterojunction with intrinsic thin layer (HIT) solar cells[29], cf. Fig. 4b, d. Very recently, the conventional HIT-cell concept was complemented by the DASH cell concept (dopant-free asymmetric heterocontacts)[30]. Bullock et al. propose alkali metal fluorides such as $LiF_x$ on intrinsic a-Si:H as electron selective heterocontacts[30] and $MoO_x$/a-Si:H(i) as hole selective heterocontacts[31] and achieve impressive conversion efficiencies. In a different approach, electron-selective contacts were realized by ALD-$TiO_2$ on ultrathin tunnel-$SiO_2$[32].

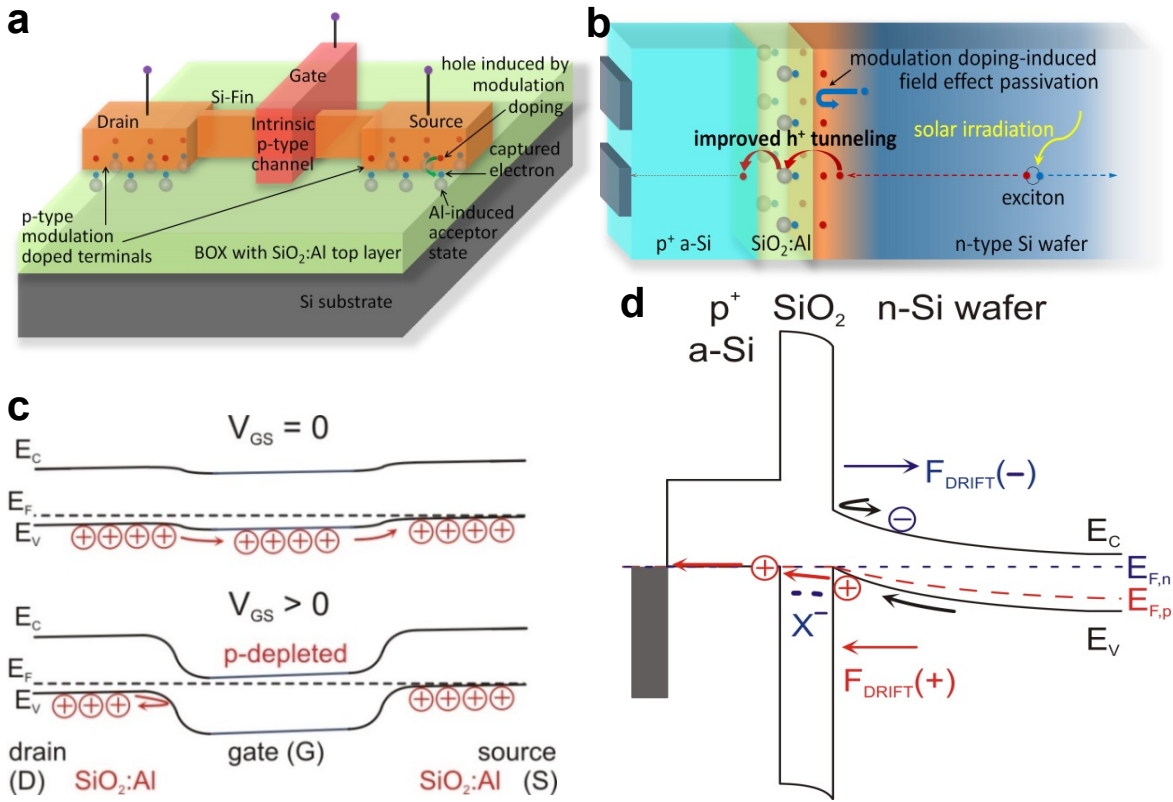

**Figure 4: Application examples for Si modulation doping. a,c,** Undoped Si fin-FETs can be provided with holes as majority carriers from Al acceptors in the buried oxide layer forming the base of the fin (**a**), eliminating any size limit due to conventional doping. The band diagram of such fin-FET (**c**) shows its working principle as hole depletion (p self-conducting) transistor. **b,d,** HIT Si solar cells can be equipped with massively enhanced hole contacts where $SiO_2$ is the optimum choice in terms of chemical bond interface passivation (**b**). Moreover, $SiO_2$ provides a much increased minority (electron) barrier while accelerating holes through a low tunnelling barrier thanks to negatively charged acceptors located about 0.5 eV below the valence band of c-Si. The band diagram of such a hole contact shows its working principle to provide much increased conversion efficiencies of HIT solar cells (**d**).

We propose that the Si modulation doping approach represents a competitive 1 technology to improve efficiencies even further: Modulation acceptors in ultrathin $SiO_2$ significantly improve hole tunnelling transport via a strongly decreased tunnelling barrier (0.5 eV vs. 4.5 eV in undoped $SiO_2$). Additionally, the acceptor states provide a negative drift field for repelling electrons as the minority carrier type, complemented by the unparalleled chemical Si surface passivation quality of $SiO_2$. A massive increase in minority carrier lifetime and ensuing ultra-low emitter saturation current density as compared to unpassivated samples is evident from Fig. 3. Currently used amorphous Si surface layers are far from featuring such properties. Interestingly, a recent theoretical screening study investigated several elements concerning their defect state energies in $SiO_2$ with respect to Si, in order to improve field-effect passivation and transport for majorities[33]. However, the energetic position derived for Al in that study differs vastly from our theoretical DFT and experimental DLTS values.

Finally, we note that Al-O monolayers in $SiO_2$ do not constitute an $Al_2O_3$ phase. While fixed negative charges and field effect passivation were also demonstrated with 30 nm ALD-$Al_2O_3$ films[34], they are not capable of improved hole tunnelling nor compatible for Si nanoscale modulation doping due to technological incompatibility and comparatively massive layer thickness.

In summary, we developed in theory and experiment a heterostructure Si-modulation doping method based on aluminium-induced acceptor states in $SiO_2$. Providing holes as majority charge carriers with such a fundamental principle represents a paradigm shift in silicon science and technology. It allows for a different strategy to define and control majority charge carriers in nanoelectronic devices and allows for passivating hole-selective contacts in high-efficiency solar cells. More generally, our modulation doping concept is transferable to other group-IV semiconductors, such as diamond or germanium, if energetically suitable combinations of impurity elements and dielectric matrices are found.

**Methods**

**Density functional theory simulations and Poisson solver.** Approximants were structurally optimised for the maximum integral over all bond energies defining the most stable configuration, using the Hartree-Fock (HF) method with 3-21G molecular orbital basis set (MO-BS)[15,35] for structural optimisations and the B3LYP hybrid DF[36,37] with 6-31G(d) MO-BS[15,38] for electronic structure calculations with the Gaussian09 software suite[39]. RMS and peak force convergence limits were 8 meV/Å and 12 meV/Å, respectively. Ultrafine integration grids and tight convergence criteria were applied to the self-consistent field routine. During all calculations, no symmetry constraints were applied to the MOs. Further accuracy evaluations can be found elsewhere[15,16]. A one-dimensional (1-D) Poisson solver for MIS structures was coded in MatLab following Nicollian and Bruce[17].

**Fabrication and characterisation of SiO$_2$:Al MOS samples.** PECVD SiO$_2$ tunnel oxides (1.5 nm) were deposited[40] onto wet-chemically cleaned 4" Czochralski P-doped Si wafers (1.6 Ωcm) in a modified Oxford Instruments PlasmaLab–FlexAL cluster. Subsequently, the wafers were transferred under vacuum into the thermal ALD chamber for deposition of Al–O monolayers via TMA and H$_2$O at 200°C (sample Al1: 1 ALD cycle, Al2: 2 ALD cycles, Al0: H$_2$O pulse only). Finally, the capping SiO$_2$ layer (10 nm) was deposited in the PECVD chamber. All wafers were RTP-annealed (1000°C, 30 s, Ar atmosphere). Electrical top- and substrate contacts (Al) were thermally evaporated and lithographically structured to fabricate MOS capacitors. DLTS and HF-CV were measured with a PhysTech FT1030 High energy resolution analysis (HERA) setup using a JANIS VPF 800 Cryostat. Minority (hole) carrier lifetime was measured as average value over the entire wafer with µW-PCD using a lifetime tester from Sinton Consulting.

## Acknowledgements

D.K. acknowledges use of Leonardi compute cluster, engineering faculty, use of Abacus compute cluster, IMDC, UNSW and funding by the 2015 UNSW Blue Sky Research Grant. D.K. and D.H. are thankful to M. Pomaska, IEK-5, RC Jülich, Germany, for µW-PCD measurements and acknowledge funding by 2012, 2014 and 2016 DAAD-Go8 joint research cooperation schemes. D.H. acknowledges the German Research Foundation (DFG) for funding (HI 1779/3-1) and thanks the IMTEK clean room team (RSC) for technical support.

## Contributions

D.K. received ideas, derived theories and conceptual applications, carried out DFT computations and coded the 1-D Poisson solver, measured and interpreted HF-CV and DLTS data, participated in sample design and drafted the manuscript. D.H. designed and processed samples, developed respective preparation techniques and recipes, participated in electrical analyses and drafted the manuscript. S.G. participated in sample processing and characterisation. M.Z. and S.S. supervised the project and provided vital resources (hardware, software, characterisation, processing). All authors revised the manuscript.

## Competing financial interests

The authors declare no competing financial interests.